\documentclass[prb,twocolumn,superscriptaddress,showkeys]{revtex4}

\usepackage{graphicx}
\usepackage{dcolumn}
\usepackage{bm}

\begin{document}

\title{\textit {Ab initio} Monte Carlo simulations for finite-temperature
properties: Application to lithium clusters and bulk liquid lithium}

\author{Sanwu Wang}
\thanks{Corresponding author. Tel.: +1 615 3432119;
              fax: +1 615 3437697.
              E-mail address: sanwu.wang@vanderbilt.edu (Sanwu Wang)}
\affiliation{Department of Physics and Astronomy,
             Vanderbilt University, Nashville, Tennessee 37235}
\affiliation{School of Computational Science and Information Technology,
             and Center for Materials Research and Technology,
             and Department of Physics, Florida State University,
             Tallahassee, Florida 32306-4120}
\author{Steven~J. Mitchell}
\affiliation{School of Computational Science and Information Technology,
             and Center for Materials Research and Technology,
             and Department of Physics, Florida State University,
             Tallahassee, Florida 32306-4120}
\affiliation{Laboratory of Inorganic Chemistry and Catalysis,
             Eindhoven University of Technology,
             5600 MB Eindhoven, The Netherlands}
\author{Per~Arne~Rikvold}
\affiliation{School of Computational Science and Information Technology,
             and Center for Materials Research and Technology,
             and Department of Physics, Florida State University,
             Tallahassee, Florida 32306-4120}


\begin{abstract}

\textit {Ab initio} Monte Carlo simulations have been performed to
determine the equilibrium properties of liquid lithium and lithium
clusters at different temperatures. First-principles density-functional
methods were employed to calculate the potential-energy change for each
proposed change of configuration, which was then accepted or rejected
according to the Metropolis Monte Carlo scheme. The resulting structural
properties are compared to data from experimental measurements and \textit
{ab initio} molecular dynamics simulations. It is shown that accurate
structural information can be obtained with \textit {ab initio} Monte
Carlo simulations at computational costs comparable to \textit {ab initio}
molecular dynamics methods. We demonstrate that \textit {ab initio} Monte
Carlo simulations for the properties of fairly large
condensed-matter systems at nonzero temperatures are feasible.

\end{abstract}


\maketitle

\section{ INTRODUCTION}

Molecular dynamics (MD) and Monte Carlo (MC) simulations have been the
major techniques for calculating finite-temperature properties of
condensed-matter systems. In particular, \textit {ab initio} MD
simulations, which combine classical molecular dynamics with
first-principles quantum theory, determine the atomic and electronic
structures of a system simultaneously and accurately, as well as the
temperature-dependent properties and time evolution of the system.
\cite{Car,Payne} \textit {Ab initio} MD methods have been widely used to
provide reliable and accurate information for various systems since Car
and Parrinello first introduced their work in 1985. \cite{Car}

On the other hand, \textit {ab initio} Monte Carlo simulations, which
employ first-principles quantum theory to calculate the potential
energy of a system at each classical MC step, have so far been used only
to a very limited extent. The majority of MC simulations for
condensed-matter systems have been performed with empirical or
semiempirical atom-atom interaction potentials, which are fitted to
available experimental data or \textit {ab initio} calculations. While the
atom-atom potentials may work well in some cases, they do not provide
reliable results in other instances. Since the quantum many-body effects
and the detailed electronic structures cannot be included in empirical or
semi-empirical potentials, first-principles theory would obviously be
desirable for providing reliable energetics. The major concern with
\textit {ab initio} Monte Carlo simulations is the high computational
demand. In particular, Monte Carlo simulations with empirical potentials
often generate millions of sampling configurations. It is a formidable
task to determine the potential energies of such a huge number of
configurations with \textit {ab initio} total-energy calculations. In
recent years, however, several groups have performed \textit {ab initio}
MC simulations for very small clusters, and they have tried to reduce the
number of sampling configurations (to typically $10^4$).
\cite{Keshari1,Truong,Asada,Akhmatskaya,Estrin,Weht,Jellinek,Watanabe,%
Bandyopadhyay,Bacelo1,Srinivas,Ishikawa}
Ishikawa \textit {et al} \cite{Keshari1,Bacelo1,Ishikawa} applied \textit
{ab initio} calculations to the Monte Carlo simulated annealing algorithm
\cite{Kirkpatrick}. They performed simulations for the Li$_6$ cluster,
small hydrogenated lithium clusters (Li$_5$H, Li$_4$H$_2$, and Li$_7$H),
and HCl(H$_2$O)$_n$ ($n = 3,4$) clusters. \cite{Keshari1,Bacelo1,Ishikawa}
Other groups \cite{Truong,Asada,Estrin,Watanabe} used the \textit {ab
initio} molecular orbital (MO) method in classical MC simulations to study
the interactions of ions and molecules with small water clusters.
Structural and thermal properties of water dimers \cite{Bandyopadhyay} and
lithium clusters (Li$_4$, Li$_5^+$, and Li$_8$)
\cite{Weht,Jellinek,Srinivas} were also investigated with \textit {ab
initio} MC simulations.

We emphasize that all the previous \textit {ab initio} MC simulations were limited
to very small clusters. Since experimental data for the structures of the
clusters at nonzero temperatures were not available, direct comparisons
of the above simulations with experiments were not possible. Therefore, the
accuracy and efficiency of \textit {ab initio} MC simulations remain to be
determined. Furthermore, the feasibility of the extension of \textit {ab
initio} MC simulations to bulk systems, which require models containing
many more atoms than the small clusters, has yet to be tested. As a first
application of \textit {ab initio} MC simulations to bulk systems, we have
chosen the liquid phase of bulk lithium. We have also investigated the
structural properties of a cluster containing 16 lithium atoms at
different temperatures. The simulated results for the lithium cluster are
physically reasonable, and the structure of liquid lithium obtained from our
\textit {ab initio} MC simulations is in good agreement with the available
experimental measurements. We also demonstrate that the computational
demands of simulating liquid lithium with our \textit {ab initio} MC
algorithms are comparable to those of \textit {ab initio} MD simulations.

The remainder of this paper is organized as follows. Section II describes   
our {\it ab initio\/} MC algorithm. Section III describes the simulation
results for a 16-atom lithium cluster (Sec.~III A) and for liquid bulk
lithium (Sec.~III B), with a discussion of general aspects and possible
improvements of our {\it ab initio\/} MC algorithm in Sec.~III C. Sec.~IV
contains our conclusion.

\section{ \textit {AB INITIO} MONTE CARLO ALGORITHM}

To simulate the cluster of 16 lithium atoms, the atoms were put in a
periodically repeated cubic supercell. The size of the supercell
($10\times10\times10$~\AA$^3$) was much larger than the concentrated
cluster (see Fig.~1.). Bulk lithium was represented by a cubic box (supercell) with
periodic boundary conditions containing 54 Li atoms. The size of the
supercell ($10.698\times10.698\times10.698$~\AA$^3$ and
$10.874\times10.874\times10.874$~\AA$^3$ for simulations at $T = 523$~K
and $T = 725$~K, respectively) was determined to match the
experimental density of liquid lithium at the given temperature.
\cite{Ohse}

The simulations were performed in canonical ensembles. The standard
Metropolis algorithm, \cite{Metropolis,Landau} which provides an efficient
approach for simulating the equilibrium properties of an atomic system at
a given temperature, was used in our \textit {ab initio} MC simulations.
Starting from a pre-selected initial configuration, the simulations
repeated the following steps. (1)  Given the current configuration
$I$ of the system, a new configuration $J$ is generated by making a small
random displacement of one randomly chosen particle (suitable for small
systems) or random displacements of all the particles simultaneously
(more efficient for systems containing a large number of particles). (2)
Once the total energies ($E_I$ and $E_J$) of these two configurations are
calculated, the acceptance probability of the new configuration $J$ is
then determined as
\begin{equation}
P(J|I) = \min\left[1, \exp\left({-(E_J-E_I) \over k_BT}\right)\right]
\;,
\end{equation}
where $T$ is the temperature and $k_B$ is Boltzmann's constant. (3) If the
configuration $J$ is accepted, it serves as the current configuration of
the next MC step, and $E_I$ is set equal to $E_J$. If the configuration
$J$ is not accepted, the configuration $I$ and its energy $E_I$ are
retained and used to start the next step. In this way, the system will
eventually reach equilibrium and evolve toward a Boltzmann distribution.
\cite{Metropolis,Landau} The resulting sequence of sampling
configurations is then used to obtain the equilibrium properties of the
system.

To calculate the total energy at each MC step, we employed an \textit {ab
initio} pseudopotential density-functional-theory (DFT) method with a
plane-wave basis set. All the \textit {ab initio} total-energy
calculations were carried out using the Vienna \textit {ab-initio}
simulation package (VASP). \cite{Kresse1,Kresse2,Kresse3} The electronic
exchange-correlation effects were treated with the generalized
gradient-corrected functionals given by Perdew and Wang.
\cite{Perdew1,Perdew2} We adopted the Vanderbilt ultrasoft pseudopotential to
replace the core electrons \cite{Vanderbilt,Kresse4} and used the
conjugate-gradient technique for performing electronic relaxations
\cite{Payne}. The total-energy calculations were conducted with 4 \textbf
{k}-points in the three-dimensional Brillouin zone with a plane-wave
energy cutoff of 120 eV. The convergence of the total energy differences
between different configurations was checked for selected configurations
with higher cutoff energies (up to 200 eV) and more \textbf {k}-points (up
to 16 \textbf {k}-points), and the convergence was found to be within a
few meV. Because we used a pseudopotential approach, the total-energy
calculations for the configurations with too small distances between
atoms, which result in core-core overlap, would give wrong total energies.
We therefore defined a hard-core distance of 2.0~{\AA} between any two Li
atoms. Trial configurations with any interatomic distance smaller than the
cutoff distance were always rejected in the MC simulations. This is
physically reasonable because such configurations would have very high
potential energies due to the strong core-core repulsion.

\section{RESULTS AND DISCUSSION}
\subsection{Small cluster}

Starting from the configuration with the atoms at ordered bcc positions
(shown in Fig. 1), we first performed our \textit {ab initio} MC
simulations for the cluster of 16 lithium atoms at temperatures of 1000~K
and 5000~K. The MC moves in the simulations for the cluster consisted of  
random displacement of one randomly selected lithium atom at each MC step,
with the maximum displacement adjusted to give an overall acceptance rate
of about 50\%. The chosen atom was displaced from its old position with
equal probability to any position inside a sphere surrounding the old
position, with the radius of the sphere being the maximum displacement. We
obtained maximum displacements of 0.75~{\AA} and 1.2~{\AA} for the
simulations at the temperatures of 1000~K and 5000~K, respectively. About
10000 MC steps were needed to reach the equilibrium state. Averages for
calculating the structural properties were taken over the next 10000
sampling configurations after equilibration. It took approximately 1
minute of CPU time to obtain the total energy of each configuration with a
single 375 MHz processor on an IBM SP3 supercomputer.

The pair correlation function $g(r)$ can be defined as \cite{Allen}
\begin{equation}
g(r) =
{2 V \over N(N-1)}
{1 \over 4{\pi}r^2}
\langle \sum_i \sum_{j<i}
\delta{(r-r_{ij})}\rangle
\;,
\label{eq:gr}
\end{equation}
where $V$ is the system volume, $N$ is the number of particles, $r_{ij}$
is the distance between a pair of particles $i$ and $j$, and $\langle  
\cdot \rangle$ signifies averaging over the sequence of simulated
configurations. The pair correlation function is proportional to the
probability of finding a pair of particles a distance $r$ apart,
normalized by the square of the particle density such that it approaches
unity in the limit of a uniform random distribution at the same density. 
It hence provides a measure of the local spatial ordering of the system.
As described in Ref.~26, we determined $g(r)$ by dividing the
range of $r$ ([2~{\AA} , 10~{\AA}]) into 400 intervals of length
0.02~{\AA} ($\Delta r$), calculating every $r_{ij}$ for a given
configuration, and recording in narrow bins $[r, r + \Delta r)$ the
frequency of occurrence of the different particle separations. In this
way a histogram was built up and finally normalized at the end of the scan
over the configurations.

\begin{figure}
\includegraphics[width=0.4\textwidth]{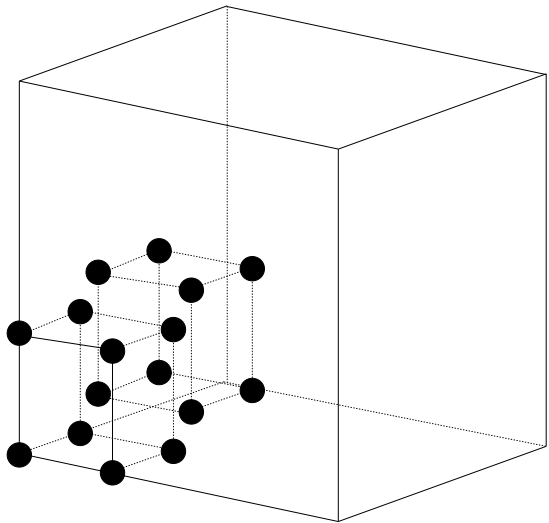}
\caption{The initial configuration of the cluster of 16 lithium atoms
         for the \textit {ab initio} MC simulations. The distance between
         two nearest-neighbor atoms is 2.96 {\AA}, the same as that
         determined by our \textit {ab initio} total-energy calculations
         for the crystalline bulk  Li. The size of the
         supercell is $10\times10\times10$ \AA$^3$.}
\end{figure}

\begin{figure}
\includegraphics{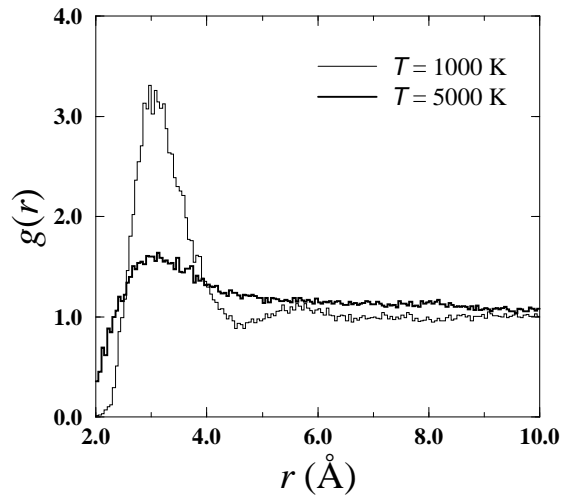}
\caption{Equilibrium pair correlation functions \textit {g}(\textit{r}) for a cluster
         of 16 lithium atoms at temperatures of 1000~K and 5000~K,
         obtained from our \textit {ab initio} Monte Carlo simulations.}
\end{figure}

Figure~2 shows the pair correlation functions for the cluster at
temperatures of 1000~K and 5000~K, respectively. It can be observed that
there are two peaks in the pair correlation function at $T = 1000$~K. This
indicates the existence of two atomic shells for the structure, and
therefore the system locally has a liquid-like structure. At a temperature
of 5000~K, however, there is no obvious second peak in the pair
correlation function, and thus the system has a gas-like structure. Given
that the boiling point of bulk Li is about 1600~K, the results from our
simulations for the cluster of lithium atoms are physically reasonable.
Note that the pair correlation function $g$($r$) at $T = 5000$~K does not
approach zero when $r$ is close to 2.0~{\AA}, indicating that the
configurations with very small separations (smaller than 2.0~{\AA})
between atoms would contribute to the ensemble averages. This reflects the
gas-like structure of the system at $T = 5000$~K because the probability
of  occurrence of configurations with very small separations between
particles is expected to be higher for a gas-like structure than for a
liquid-like structure.

\subsection{Bulk system}

Following the simulations for the lithium cluster, we carried out \textit
{ab initio} MC simulations for the liquid phase of bulk lithium. In
contrast to the simulations for the lithium cluster, the Monte Carlo moves
in the simulations for liquid lithium consisted of random displacements of
\textit {all} the atoms in the supercell simultaneously. The other aspects
of the MC moves were the same as in the simulations for the lithium
cluster. In order to achieve an overall acceptance rate of roughly 50\%,
we found that the maximum displacements at temperatures of 523~K and 725~K
were 0.05~{\AA} and 0.06~{\AA}, respectively. We started our simulations
at $T = 725$~K from an initial configuration with all the Li atoms at
their bulk bcc positions. Roughly 8000 MC steps were required to reach
equilibrium. Since the initial configuration was a perfect crystalline
structure, far away from the equilibrium structure of the liquid phase of bulk
lithium, it required relatively long MC runs to reach equilibrium.
In contrast, the initial configuration for the simulations at $T =$
523~K was chosen as one of the equilibrium sampling configurations
obtained in the simulations at $T = 725$~K. A reduced number of MC steps
(5000 MC steps) was then required for the equilibration period. After
equilibration at either temperature, we generated 5000 sampling
configurations for averages.

Our simulations were carried out in parallel on four 375-MHz processors of
an IBM SP3 supercomputer. The total-energy calculation at each MC step
required approximately 150 seconds of CPU time. On average, the electronic
relaxation converged within seven self-consistency steps. The software
(VASP) that we used for the total-energy calculations contains
computations for the forces after the electronic relaxation converges.
Such computations are not needed in \textit {ab initio} MC simulations.
Should VASP be modified to exclude the force calculations, the
computational costs would therefore be decreased by approximatively 10\% 
-- 15\%.

For comparison, we also performed \textit {ab initio} MD simulations
\cite{Kresse1,Kresse2,Kresse3} for liquid lithium. Canonical ensembles  
were generated with the algorithm of Nos{\'e}, \cite{Nose} and the time   
step was set to 1 fs. The initial configuration for the \textit {ab
initio} MD simulations at both $T =$ 725~K and $T =$ 523~K was the same
perfect bcc structure as that for the \textit {ab initio} MC simulations
at $T =$ 725~K. The equilibration periods extended over 5000 and 3000 time
steps at $T =$ 523~K and $T =$ 725~K, respectively. Ensemble averages were
calculated from the sampling configurations of the next 5000 steps. Each
step required about 400 seconds of CPU time on an IBM SP3 supercomputer   
(in parallel on 4 processors), and the corresponding wall-clock time was
roughly 100 seconds.

\begin{figure}
\includegraphics{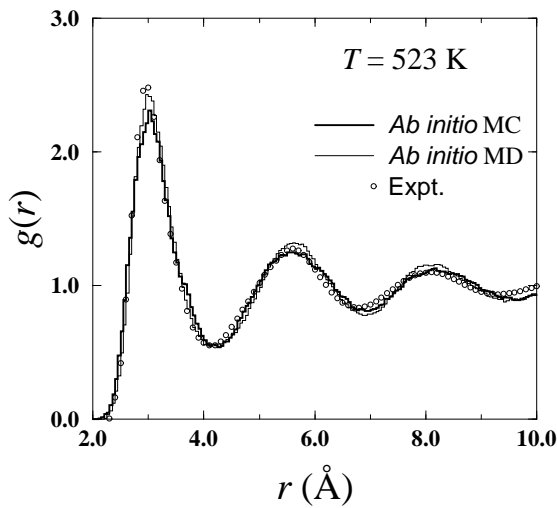}
\caption{Equilibrium pair pair correlation function \textit {g}(\textit{r}) for liquid
         lithium at a temperature of 523 K, calculated from \textit {ab
         initio} Monte Carlo and \textit {ab initio} molecular dynamics
         simulations (histograms). Also shown are experimental data
         measured by X-ray diffraction (circles). \cite{Waseda}}
\end{figure}

\begin{figure}
\includegraphics{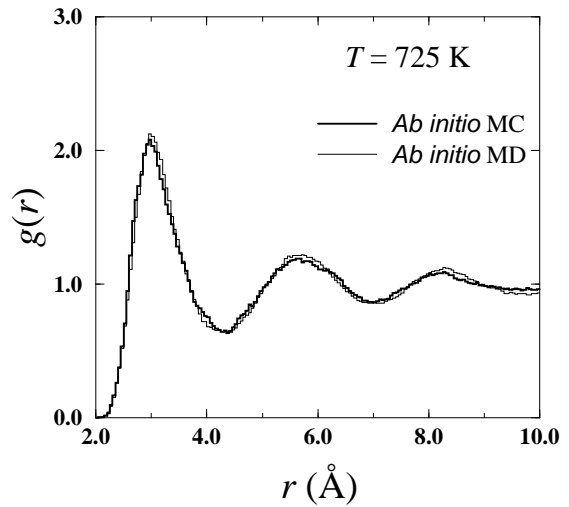}
\caption{Equilibrium pair correlation function \textit {g}(\textit{r}) for the liquid
         phase of lithium at a temperature of 725 K, obtained from \textit
         {ab initio} MC and \textit {ab initio} MD simulations.}
\end{figure}

The pair correlation functions for liquid lithium at temperatures of 523~K
and 725~K, obtained from our \textit {ab initio} MC and \textit {ab
initio} MD simulations are shown in Fig. 3 and Fig. 4, respectively. For
comparison, the experimental data for liquid lithium at $T = 523$~K
obtained by X-ray diffraction \cite{Waseda} are also shown in Fig. 3. We
observe that the agreement between \textit {ab initio} MC, \textit {ab
initio} MD, and experiment is very good. In particular, the three peaks
determined from our \textit {ab initio} MC simulations are essentially
located at the experimental positions. The intensities of the peaks
obtained from the \textit {ab initio} MC simulations are also basically in
agreement with those from both the \textit {ab initio} MD simulations and
the X-ray data. The small differences of the intensities of the peaks
between the theoretical results and the experimental data may be due to
the small size of our supercells. The experimentally observed temperature
effects, \cite{Olbrich} that is,  broadening of the peaks and slight
shifts to higher positions of the second and third peaks at the higher
temperature are also well reproduced by our \textit {ab initio} MC
simulations (see Fig.~3 and Fig.~4).

\subsection{General discussion}

Finally, we discuss some general aspects and possible improvements of our
\textit {ab initio} MC algorithm. The computational costs of \textit {ab
initio} MC simulations are dominated by the \textit {ab initio}
total-energy calculations. Since efficient iterative schemes for \textit
{ab initio} total-energy calculations with a plane-wave basis set have
achieved order $N^2$ ($N$ is the number of the atoms in the supercell)
scaling for systems containing up to 1000 electrons, \cite{Kresse3} the
cost of simulations with our \textit {ab initio} MC algorithm would
increase essentially as the square of the number of atoms.

While the computational costs are comparable, our \textit {ab initio} MC
algorithm is not any faster than \textit {ab initio} MD simulations. In 
particular, a larger number of steps are required to reach equilibrium in
our \textit {ab initio} MC simulations for bulk lithium than in the
corresponding \textit {ab initio} MD simulations. For the lithium systems
that we have investigated, our \textit {ab initio} MC algorithm produces
the required sampling configurations at a speed similar to the \textit {ab
initio} MD simulations. In this respect, our \textit {ab initio} MC
simulations are as efficient as \textit {ab initio} MD simulations.

In some cases, like simulations for strongly covalent systems and
situations involving volume changes, however, our \textit {ab initio} MC
algorithm may not be efficient due to the low acceptance rate of the
traditional MC moves. In such situations, therefore, our \textit {ab
initio} MC algorithm needs to be improved. A cluster MC move scheme
proposed by Lee and Swendsen \cite{Lee} may be used for this purpose. We
have tested the cluster MC move scheme for the bulk lithium systems that
we investigated with traditional MC moves as described in Sec. III. We
found that the use of cluster MC moves in \textit {ab initio} MC
simulations provided almost the same performance as the traditional MC
moves, indicating that both kinds of MC moves might work equally well for
metallic systems. It is also possible that the relative performance of
different kinds of MC moves somewhat depend on the number of particles
included in the supercell. It would be interesting to see if the cluster
MC move scheme would be more efficient than the traditional MC approach in
simulations for systems containing large numbers (for example, hundreds)
of particles. In the cases where the boundary conditions or/and the
volume of the simulated cells change, which result in a low acceptance
rate with the traditional MC moves, the cluster MC move scheme has been
shown to increase the acceptance rate significantly. \cite{Lee} We also
expect that the cluster MC move scheme would be more efficient than the
traditional MC move method in simulations for systems containing strong
covalent bonds. Another approach, a hybrid of MD and MC simulation in
which trial MC moves are generated with classical MD methods, may also
enhance the efficiency of \textit {ab initio} MC simulations. \cite{Duane}
Other techniques aiming at sampling important configurations in the
configuration space include the J-walking procedure proposed by Frantz et
al. \cite{Jellinek,Srinivas,Frantz} and a big-move method suggested by
Akhmatskaya et al. \cite{Akhmatskaya}.

\section{ CONCLUSION}

In conclusion, we have shown that the use of \textit {ab initio}
calculations in classical Monte Carlo simulations provides structures of
liquid lithium in agreement with experimental measurements and \textit {ab
initio} MD simulations. Our work also demonstrates that \textit {ab
initio} MC simulations for condensed-matter systems with up to several
tens of atoms (and possibly with more than a hundred atoms) are feasible
and may provide an alternative way to investigate finite-temperature
properties of such systems.

\acknowledgments

We thank G. Brown, Y.Z. Cao, Z. Liu, M.A. Novotny, and H.K. Lee for
helpful discussions. This work was supported by the National Science
Foundation under grants DMR-9981815, DMR-0111841, and DMR-0240078,
and by Florida State University through the Center for Materials Research
and Technology and the School of Computational Science and Technology. The
calculations were performed on IBM SP3 computers at Florida State
University and the National Energy Research Scientific Computing Center,
which is supported by the U.S. Department of Energy. 

\bibliography{condmat_swwang4r}

\end{document}